\documentclass[11pt,a4paper]{article}
\usepackage[T1]{fontenc}
\usepackage[utf8]{inputenc}
\usepackage{authblk}

\usepackage{moreverb}

\usepackage{amsmath} 
\usepackage{amssymb} 
\usepackage{subcaption}
\usepackage{mathrsfs} 
\usepackage{fancyhdr} 
\usepackage{graphicx} 
\usepackage[T1]{fontenc} 
\usepackage[utf8]{inputenc} 
\usepackage{times} 
\usepackage{eurosym} 
\usepackage{color} 
\usepackage{multicol} 
\usepackage{ulem} 
\usepackage{makeidx}
 \usepackage{ulem}

\makeatletter
\def\@xfootnote[#1]{%
  \protected@xdef\@thefnmark{#1}%
  \@footnotemark\@footnotetext}
\makeatother

\usepackage{caption}
\usepackage[]{natbib}
\usepackage[colorlinks,citecolor=blue]{hyperref}

\date{}

\begin{document}


\begin{center}
{\huge
		{Item response models for the longitudinal analysis of health-related quality of life in cancer clinical trials}
}
\bigskip
\bigskip

\large{Antoine Barbieri}\footnote[$\star$]{Corresponding author: Antoine.Barbieri@umontpellier.fr}$^{,}$\footnote[1]{Université de Montpellier, Montpellier, France}$^{,}$\footnote[2]{Institut régional du Cancer Montpellier (ICM) - Val d'Aurelle, Biometrics Unit, Montpellier, France}$^{,}$\footnote[3]{Institut Montpelliérain Alexander Grothendieck (IMAG), Montpellier, France}, \large{Jean Peyhardi}\textcolor{red}{$^{1}$}$^{,}$\footnote[4]{CIRAD, AGAP and Inria, Virtual Plants, France}$^{,}$\footnote[5]{Institut de génomique fonctionnelle, Montpellier, France}, \large{Thierry Conroy}\footnote[6]{National Quality of Life in Oncology Platform, France}$^{,}$\footnote[7]{Institut de Cancérologie de Lorraine, Nancy, France}, \\ \large{Sophie Gourgou}\textcolor{red}{$^{2}$}, \large{Christian Lavergne}\textcolor{red}{$^{3}$}$^{,}$\footnote[8]{Université Paul-Valéry Montpellier 3, Montpellier, France} and \large{Caroline Mollevi}\textcolor{red}{$^{2}$}$^{,}$\footnote[9]{IRCM, Institut de Recherche en Cancérologie de Montpellier, Montpellier, France}$^{,}$\footnote[10]{INSERM, U1194, Montpellier, France} \\ 
%
\end{center}

\bigskip

\begin{abstract}
Statistical research regarding health-related quality of life (HRQoL) is a major challenge to better evaluate the impact of the treatments on their everyday life and to improve patients' care. 
Among the models that are used for the longitudinal analysis of HRQoL, we focused on the mixed models from the item response theory to analyze directly the raw data from questionnaires. 
Using a recent classification of generalized linear models for categorical data, we discussed about a conceptual selection of these models for the longitudinal analysis of HRQoL in cancer clinical trials. 
Through methodological and practical arguments, the adjacent and cumulative models seem particularly suitable for this {context}. Specially in cancer clinical trials and for the comparison between two groups, the cumulative models has the advantage of providing intuitive illustrations of results.
To complete the comparison studies already performed in literature, a simulation study based on random part of the mixed models is then carried out to compare the linear mixed model classically used to the discussed item response models. As expected, the sensitivity of item response models to detect random effect with lower variance is better than the linear mixed model sensitivity.
Finally, a longitudinal analysis of HRQoL data from cancer clinical trial is carried out using an item response cumulative model.
\end{abstract}

\textit{Keywords:}{ Item response theory; Mixed models; Ordinal categorical data; Longitudinal analysis; Health-related quality of life.}


\section{Introduction}
Endpoints refer to biological and clinical measures to assess the efficiency of new therapeutic strategies. The overall survival endpoint is the gold standard to show a clinical benefit of these strategies and treatments. Therapeutic treatments being more efficient and increasing the patients' lifetime, the overall survival endpoint may become insufficient to show a significant difference between two treatments. It is then necessary to consider a longer follow-up or a larger cohort of patients to have a sufficient number of events and a good statistical power \citep{fiteni_endpoints_2014}, both representing considerable costs. 
Thus, to conclude to the benefit of a new treatment, other endpoints have emerged and the health-related quality of life (HRQoL) is currently one of the most important. In cancer clinical trials, the patient-reported outcomes are increasingly used to analyze a clinical benefit for medical decision-making \citep{fiteni_endpoints_2014}. 
The HRQoL endpoint may seem more pertinent to show the interest of a new therapy in some cases such as the palliative or geriatric situations. However, there are conceptual and methodological brakes underlying to the concept and the assessment of HRQoL. 
{Indeed}, HRQoL is a multidimensional concept regarding the physical, psychological and social functions as well as symptoms associated with the disease and treatments. Another conceptual brake is the subjectivity of its measurement. Indeed, patients report their feelings about their HRQoL thanks to self-reported questionnaires. Both arguments preclude the use of HRQoL as sole primary endpoint in clinical trials.

In oncology, HRQoL is assessed using a general questionnaire for a set of different cancers, and an additional specific questionnaire associated with each type of cancer \citep{aaronson_european_1993,cella_functional_1993}. 
Each questionnaire decomposes the HRQoL to measure several under-concepts (dimensions of HRQoL) which themselves comprise one or several items. The items are built on the Likert scales in which the response variable is ordinal. 
Thus, considering several items for a given dimension, HRQoL data are composed of multiple ordinal responses. 
Also, the questionnaires are filled by the subjects themselves, and collected at different times defined in the trial protocol (usually at inclusion, during treatment and follow-up). These repeated measures are used to assess the evolution of the subject's HRQoL over time. 
In Europe, these questionnaires are developed and validated by the European organization for research and treatment of cancer (EORTC). The standard questionnaire currently used in oncology is the EORTC Quality of Life Questionnaire - Core 30 (EORTC QLQ-C30) \citep{aaronson_european_1993}, together with the scoring procedure proposed by the EORTC \citep{fayers_eortc_2001}. The score is then calculated for each dimension and for each subject, corresponding to the average of the item responses for a single dimension, and expressed on a scale ranging from 0 to 100. The interpretation is such that high functional scores reflect good functional capacities and a good HRQoL level, and conversely, high symptomatic scores represent strong symptoms and point out difficulties. 
The use of scoring procedures is common in practice because the statistical methods for quantitative variables are more powerful and easier to implement and interpret \citep{grilli_multilevel_2011}. But, in a Likert scale, the gap which separates each adjacent category of response ("not at all", "a little", "quite a bit" and "very much") may not be the same, and the HRQoL score calculation does not take into account this characteristic.
{Another drawback in the score use is that subjects could have different item outcomes and obtain the same score. In this situation, the score does not make a distinction between these subjects \citep{gorter_why_2015}.}

The longitudinal statistical models classically used in oncology are performed on the summary score through using the linear mixed models (LMM) or time-to-event models\citep{anota_comparison_2014}. In the LMM, the variable associated with the HRQoL score is considered as a Gaussian variable while it presents the characteristics of an ordinal variable, being non-continuous and bounded. These models allow to take into account the correlation introduced by repeated measurements on the same patient (collection of the HRQoL questionnaires over time) and different covariates such as time, treatment group, age... However, the use of the LMM for HRQoL analysis is scientifically questionable given the characteristics of the score. 
Furthermore, many symptomatic dimensions are composed of only one item, the HRQoL score has exactly the same properties than ordinal categorical data, and using the LMM is not appropriated. Thus, if a ceiling or floor effect is observed, the categorical feature is even more marked when one of the two extreme categories is over-represented. 
The second approach for the longitudinal analysis of HRQoL is based on the time-to-event models: the time-to-deterioration (TTD) and the time-until-definitive-deterioration (TUDD) \citep{hamidou_time_2011}. Survival approaches are often used and thus well-known in the oncology, and are appreciated for their easiness to interpret result and their good understanding by clinicians.
In these models, an event is classically defined by the (definitive or not) deterioration of the HRQoL score between baseline and a follow-up time, given a minimally clinically important difference \citep{anota_time_2013}. The lack of homogeneity of the methods used for the HRQoL data analyses in different oncology clinical trials is also a real obstacle to the comparison of results. Indeed, the LMM and TUDD approaches show results which may sometimes seem contradictory, and with different interpretations, but they may also be complementary. An example can be taken comparing two similar cancer clinical trials investigating the effect of bevacizumab. In the first trial \citep{chinot_bevacizumab_2014}, HRQoL analysis through TUDD showed that the bevacizumab group had a later deterioration of HRQoL compared with patients in the standard group. Conversely, in the second trial \citep{gilbert_randomized_2014}, HRQoL analysis using the LMM showed a worse HRQoL overtime in the bevacizumab group. 

Interest in the HRQoL endpoint is growing rapidly in cancer clinical trials and it is essential to {use} a suitable methodology to analyze HRQoL data, taking into account the data properties (repeated measurements of the multiple ordinal responses). 
In our study, we first focused on the different and most adapted models to analyze HRQoL from raw data, \textit{i.e.} directly on the item outcomes. Studies on psychometric properties from questionnaires such as the one used for HRQoL have been ongoing for a long time \citep{edelen_applying_2007,jafari_using_2012}, known as the item response theory (IRT). The IRT models link the individual's item responses and the latent variable which represents the studied HRQoL concept. 
They can be seen as generalized linear mixed models (GLMM) for ordinal responses with a particular parameterization of the linear predictor. The interest for this kind of model to analyze the data, including the longitudinal analyzes, is increasing \citep{titman_item_2013,hardouin_power_2015,gorter_why_2015,santos_multidimensional_2016}. 
However, to our knowledge, there is no work that discusses of the choice of one of the different IRT models over the others, and specially for HRQoL longitudinal analysis. 
First, we propose in section 2 a conceptual selection of these models through practical and methodological arguments. 
In this selection, we replace IRT models in the GLMM framework using the new specification of generalized linear models (GLM) for categorical responses, proposed by \citet{peyhardi_new_2015}.
Then, to complete the comparisons already done between the IRT models and the LMM on their capacity to detect fixed effects, we focus in section 3 on the sensibility of these models to detect the random effects through a simulation study. 
Finally, section 4 presents an application of the chosen IRT model on real data from a multicenter randomized phase III clinical trial in first-line metastatic pancreatic cancer patients \citep{conroy_folfirinox_2011}.

\section{Conceptual selection of IRT models}

This section concerns a conceptual selection of IRT models for the longitudinal analysis of HRQoL in cancer clinical trials.
HRQoL raw data are repeated measurements of ordinal multiple responses. 
The GLMM for ordinal responses seem well suitable to analyze this kind of data. The use of random effects takes into account the inter-patient variability and the correlation between the repeated measurements for each single patient.
The IRT models have been increasingly used to analyze health data deriving from self-questionnaires made of polytomous responses \citep{hardouin_spatio-temporal_2012,anota_comparison_2014,barbieri_applying_2015}. 
These models turn out to be GLMM for polytomous data with a specific parameterization of the linear predictor taking into account the multiple outcomes.  
For ordinal responses, three families of regression models are described: the families of adjacent models \citep{masters_rasch_1982,agresti_analysis_2010}, cumulative models \citep{samejima_estimation_1968,mccullagh_regression_1980}, and sequential models \citep{tutz_sequential_1990,fahrmeir_multivariate_2001}.
Many IRT models are proposed for the analysis of this kind of data, often with no explanation regarding the choice of one model over another.

In this section, we use the new specification of the GLM for categorical data proposed by \citet{peyhardi_new_2015} to talk about the relevance of the models adopted in the context of the HRQoL longitudinal analysis in cancer clinical trials.
Each model is defined according to three components \textit{(r,F,Z)}: the ratio of probabilities $(r)$, the cumulative distribution function $(F)$, and the parameterization of the linear predictor determined by the design matrix $(Z)$.
For the GLMM framework, we extended this new specification to the quadruplet \textit{(r,F,Z,U)} with $Z$ the design matrix of fixed effects and $U$ the design matrix of random effects.
The relationship between these components is determined by $R=\mathcal{F}(Z\beta+U\xi)$.
Given the linear predictor $\boldsymbol{\eta}=Z\beta+U\xi$ and $\boldsymbol{\pi}_{iv}^{(j)}=(\pi_{iv0}^{(j)},\ldots,\pi_{iv,M-1}^{(j)})$ the truncated vector\footnote{This truncated vector is sufficient to characterize the categorical distribution since $\pi_{ivM}^{(j)}=1-\sum^{M-1}_{c=0}\pi_{ivc}^{(j)}$ ;} of conditional probabilities with $\pi_{ivm}^{(j)}=\Pr\left(Y_{iv}^{(j)}=m\vert\xi_i\right)$ the conditional probability that subject $i(i=1,\ldots,n)$ selects the category $m\in\left\{0,\ldots,M_j\right\}$ for item $j(j=1,\ldots,J)$ at visit $v(v=1,\ldots,V)$ given individual random effect, we defined:
\begin{itemize}
\item $R=\left\{r_m\left(\boldsymbol{\pi}_{iv}^{(j)}\right)\right\}_{i,j,v,m}$ ; 
\item $\mathcal{F}\left\{\left({\eta}_{ivm}^{(j)}\right)_{i,j,v,m}\right\}=\left\{F\left({\eta}_{ivm}^{(j)}\right)\right\}_{i,j,v,m}$ where $\boldsymbol{\eta}=\left({\eta}_{ivm}^{(j)}\right)_{i,j,v,m}$.
\end{itemize}

After a presentation of the IRT parameterization used concerning the linear predictor, we compare different polytomous IRT models on the basis of the link function (ratio of probabilities and the cumulative distribution function (CdF)) using methodological and practical arguments.

\subsection{The IRT parameterization of the linear predictor}

The IRT probabilistic models emerged following the works of Georg Rasch \citep{rasch_probabilistic_1960} on dichotomous responses, and were then extended to ordinal responses. 
Considering the three families of adjacent, cumulative and sequential models, there are three associated famous IRT models \citep{boeck_explanatory_2004,bacci_class_2014}, respectively the graded response model \citep{samejima_estimation_1968}, the (generalized) partial credit model \citep{masters_rasch_1982,muraki_generalized_1992}, and the sequential model \citep{tutz_sequential_1990}.
These models link the individual's item responses to the unidimensional latent variable which represents a concept not directly measurable. 
In oncology framework, the concept is HRQoL relatively to one specific HRQoL dimension. 

From the IRT, the specific parameterization of the linear predictor $\eta_{im}^{(j)}$ is built into two parts: the individual part and the item part. 
The best-known way is to consider the following decomposition:
\begin{equation}\label{eq_classical_eta}
\eta_{im}^{(j)}=\alpha_j\left(\theta_i-\delta_{jm}\right),
\end{equation}
where $\theta_i$ is associated with an unidimensional random variable (currently assuming to be standard normal for identifiability) representing the latent value for the subject $i$, $\delta_{jm}$ and $\alpha_j$ being the item parameters. 
Generally called difficulty parameter, $\delta_{jm}$ is the intercept (or threshold) associated with the item $j$ for the category $m\in\left\{1,\ldots,M_j\right\}$. 
The parameter $\alpha_j$ is called the discrimination parameter of item $j$, and represents the sensitivity of each response probability according to the value of the latent trait. 
Indeed, the more the discrimination parameter value is high, the more the item allows well discriminating two individuals with a close latent trait value.
However, the predictor is no longer linear for IRT models using discrimination parameters because it includes a product of parameters. Thus, these models do not belong to the class of GLMM \citep{liu_mixed-effects_2006}.

In oncology, HRQoL analysis is carried out using IRT models which do not include the discrimination parameters (fixed to one for all items). Thus, these IRT models are within the class of GLMM.
Concerning the longitudinal analysis, several studies proposed to extend some IRT models using the linear decomposition of the latent variable $\theta$ with fixed and random effects \citep{hardouin_spatio-temporal_2012,verhagen_longitudinal_2013,huber_mathematical_2013}:
\begin{equation}
\theta_{iv}=x_{iv}'{\beta}+u_v'{\xi}_i,
\label{eq_thetadecomp}
\end{equation}
with the vector ${\beta}$ is associated with the fixed effects, the vector ${\xi_i}$ with the subject-specific random effects and the index $v$ the current visit.
In the equation \eqref{eq_thetadecomp},
$\theta_{iv}$ is thus the estimation of latent process at the visit $v$.

\subsection{The probability ratio: structure of the models}

The linear predictor is not directly related to the response probability but to a particular transformation ratio. 
The choice of the ratio is related to the nature of response from the ordering assumption among categories. Thus, reference ratio \citep{peyhardi_new_2015} for nominal response is excluded in this work because the HRQoL responses are ordinal.
Let's consider the simple situation from GLM with one item with $(M+1)$ response categories.
The three model families for ordinal data are distinguished by the choice of the ratio of probabilities $\boldsymbol{r}\left(\boldsymbol{\pi}\right)=\left(r_0\left(\boldsymbol{\pi}\right),\ldots,r_{M-1}\left(\boldsymbol{\pi}\right)\right)$. 
Each model is summarized by $M$ equations $\left\{r_{m}\left(\boldsymbol{\pi}\right)=F\left(\eta_{m}^{\star}\right)\right\}_{m=0,\ldots,M-1}$ with $\eta_{m}^{\star}=\delta_m-\theta$, highlighting the decomposition of the link function which is determined through the ratio of probabilities and the CdF. 
Indeed, we may distinguish different ratios of probabilities for these different families, respectively, 
for the cumulative models, 
\begin{eqnarray}
r_{m}\left(\boldsymbol{\pi}\right) = \pi_{0}+\ldots +\pi_{m},\quad m=0,\ldots,M-1; \label{ratio_cum}
\end{eqnarray}
for the adjacent models,
\begin{eqnarray}
r_{m}\left(\boldsymbol{\pi}\right)=\frac{\pi_{m}}{\pi_{m}+\pi_{m+1}},\quad m=0,\ldots,M-1; \label{ratio_adj}
\end{eqnarray}
and, for the sequential models,
\begin{eqnarray}
r_{m}\left(\boldsymbol{\pi}\right)=\frac{\pi_{m}}{\pi_{m}+\ldots +\pi_{M}},\quad m=0,\ldots,M-1.\nonumber
\end{eqnarray}

In the IRT, adjacent and cumulative models are usually presented given the reverse permutation \citep{samejima_estimation_1968,masters_rasch_1982,bacci_class_2014}. 
This permutation is defined as the reversal of category order \citep{mccullagh_regression_1980}. Assuming that the considered CdF is symmetric (i.e. the corresponding probability density function is symmetric about the y-axis), these models are invariant under this permutation \citep{peyhardi_new_2015}.
For our application context, this is as an advantage for result interpretation. A lower item-response category reflects a lower level of the symptomatic dimensions whereas it represents a higher level of capacity for the functional dimensions. The reverse permutation for the functional dimensions makes it easier and intuitive for clinicians to present their results. This allows the homogenization of the result interpretation as it is done in the scoring procedure proposed by the EORTC (for functional dimensions, the score scale is reversed compared with the item responses categories order) \citep{fayers_eortc_2001}. 
However, sequential models correspond to process ordering and reversing the process may change its nature. 
These models are not reversible (i.e. no invariant under the reverse permutation).
Thus, sequential models will not be used and only the adjacent and cumulative models which correspond to scale ordering as used for HRQoL measurements, will be consider.

The cumulative models also have additional properties \citep{mccullagh_regression_1980}, including that they are invariant when successive categories are gathered. Thus, if one category is not observed, it can be combined with its successive categories without changing the model.
Another advantage of the cumulative models is their interpretation through a continuous latent variable. 
Indeed, the continuous latent variable $\tilde{Y}$ underlying the model exists and a direct link with the response variable $Y$ through the thresholds presumed to be strictly increasing ($-\infty=\delta_{0}<\delta_{1}<\ldots<\delta_{M}<\delta_{M+1}=+\infty$) is such as:
\begin{eqnarray}
\left\{Y=m\right\} & \mbox{if} & \left\{\delta_{m}<\tilde{Y}\leq\delta_{m+1}\right\},\quad m=0,\ldots,M\;, \nonumber
\end{eqnarray} 
where $\tilde{Y}=\theta+\varepsilon$ and $\varepsilon$ is the error term distributed following the CdF.  
Here, the latent variable $\tilde{Y}$ represents HRQoL and its interpretation is then equivalent to the interpretation of the response variable using a LMM.

However, an advantage of the adjacent models is that there is no constraint affecting the model estimation. 
Nonetheless, the cumulative models have to respect constraints, which can make difficult the model estimation, particularly in the case of non-proportional design of linear predictor \citep{peyhardi_new_2015}.
For the proportional design, a common slope ($\theta$) is considered for all categories, else the slope is dependent of the category ($\theta_m$). 
Let the simple parameterization of the linear predictor $\eta_m=\theta_m-\delta_m$ for $m\in\left\{1,\ldots,M\right\}$ where $\theta_m$ and $\delta_m$ are the slope and the intercept associated with the category $m$, respectively. 
Considering proportional design ($\theta=\theta_1 =\ldots=\theta_{M}$), the cumulative models refer to the principle of thresholds \citep{mccullagh_regression_1980,hedeker_random-effects_1994} with the constraint they have to be strictly increasing such as $-\infty<\delta_{1}<\ldots<\delta_{M}<+\infty$. 
Considering the non-odd proportional models, the constraint then becomes $-\infty<\eta_M<\ldots<\eta_{1}<+\infty$ which is more difficult to verify. 
For the longitudinal analysis of HRQoL data in oncology, the proportional design is considered and to verify the constraint only on the threshold is easier.

Table \ref{table_recap} summarizes some characteristics of these three families of models which are important for the longitudinal analysis of HRQoL in cancer clinical trials. 
In this context, a proportional design of the linear predictor is preferred. Under this parameterization, there is few difficulties to respect the cumulative models constraints and to estimate them. 
The adjacent models seem more flexible than cumulative models because they are always define for all linear predictor. But, their interpretation of the results using the cumulative model is more intuitive than adjacent models. 

\begin{table}[t!]
\caption{Summary of the characteristics for the three model families \label{table_recap}}
\centering
\scalebox{1}{
\begin{tabular}{lccc}
\hline
 & \multicolumn{3}{c}{Models}\\
\cline{2-4}
Characteristics &  Adjacent & Cumulative & Sequential \\
\hline
Reversibility & $yes$ & $yes$ & $no$ \\
Interpretation using the latent variable & $no$ & $yes$ & $yes$ \\
Always defined  & $yes$ & $yes(no^{1})$ & $yes$ \\
\hline
\multicolumn{2}{l}{\footnotesize{$^{1}$: for some non proportional design models}} & \multicolumn{2}{r}{}\\
\end{tabular} 
}
\end{table}

\subsection{The cumulative distribution function}

The latest component discussed in the IRT model selection is the CdF. 
Each model probability can be defined with any CdF.
As commonly seen in the IRT models, and thank to the reversibility property, the adjacent and cumulative models are used in descending order.
For the cumulative model, the probabilities are defined from the equation \eqref{ratio_cum} and given the $F$ as:
\begin{equation}\label{cum_cdf}
  \left\{
      \begin{array}{ccl}
       \pi_{0} & = & 1-  F\left(\eta_{1}\right)\\
\pi_{m} & = & F\left(\eta_{m}\right) - F\left(\eta_{m+1}\right) ,\quad m=1,M-1\\
\pi_{_M} & = & F\left(\eta_{_M}\right)      
    \end{array}
    \right. .
\end{equation}
Then, general expression of the sequential model whatever the CdF used is such that 
$$\pi_m=F(\eta_m)\prod_{k=1}^{m-1}\left\{1-F(\eta_{k})\right\} ,$$
where $m=1,\ldots,M$ and $\prod_{k=1}^0\{.\} =1$ \citep{fahrmeir_multivariate_2001}.

Such general equations have never been presented for adjacent models, only described with the logistic CdF \citep{masters_rasch_1982,muraki_generalized_1992,fahrmeir_multivariate_2001,agresti_analysis_2010,hardouin_spatio-temporal_2012,anota_comparison_2014}. 
However, the different response probabilities can be presented from the adjacent ratio and according to a general CdF ($F$):
\begin{equation} \label{adj_cdf}
  \left\{
      \begin{aligned}
       \pi_0 & = & \frac{1}{1+\sum_{m=1}^{M}\prod_{k=1}^{m}\left(\frac{F(\eta_k)}{1-F(\eta_k)}\right)}\\
\pi_m & = & \frac{\prod_{k=1}^{m}\left(\frac{F(\eta_k)}{1-F(\eta_k)}\right)}{1+\sum_{m=1}^{M}\prod_{k=1}^{m}\left(\frac{F(\eta_k)}{1-F(\eta_k)}\right)}&,\quad m=1,\ldots,M
      \end{aligned}
    \right.
\end{equation}

The CdF choice is especially used to best fit the data.
Let's four CdF from two different kinds: the most commonly used symmetric distributions, the logistic and Gaussian distributions, and the two asymmetric distributions, the Gumbel min and Gumbel max distributions. 
The two later distributions are respectively defined by $F(\eta)=exp(-exp(-\eta))$ for the Gumbel max distribution and by $F(\eta)=1-exp(-exp(\eta))$ for the Gumbel min distribution. 

\begin{figure}[tbh!]
\centering
\begin{subfigure}{0.48\textwidth}
  \centering
  \includegraphics[width=\textwidth]{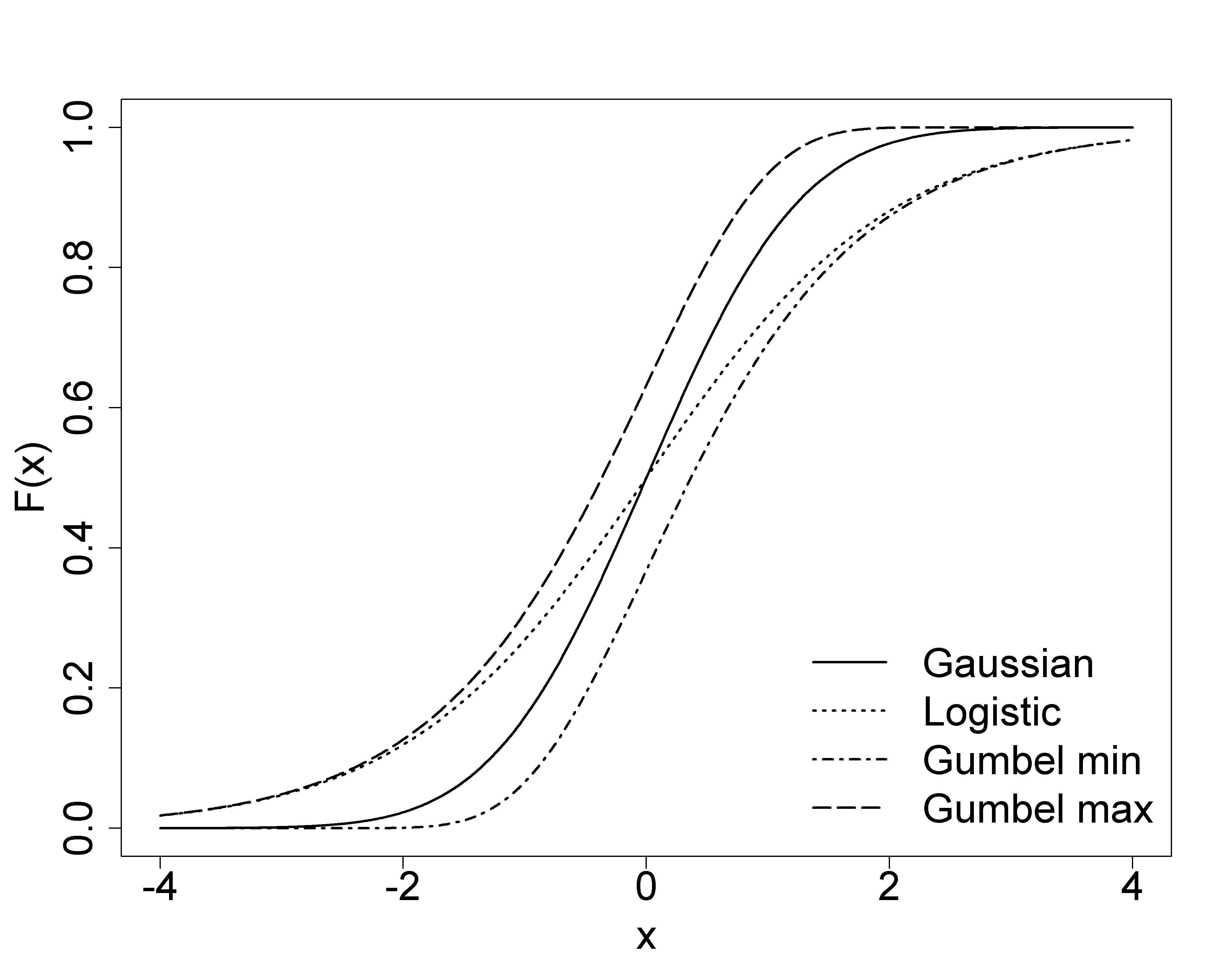}
  \caption{CdF}
  \label{fig_cdf}
\end{subfigure}
     ~ 
\begin{subfigure}{0.48\textwidth}
  \centering
  \includegraphics[width=\textwidth]{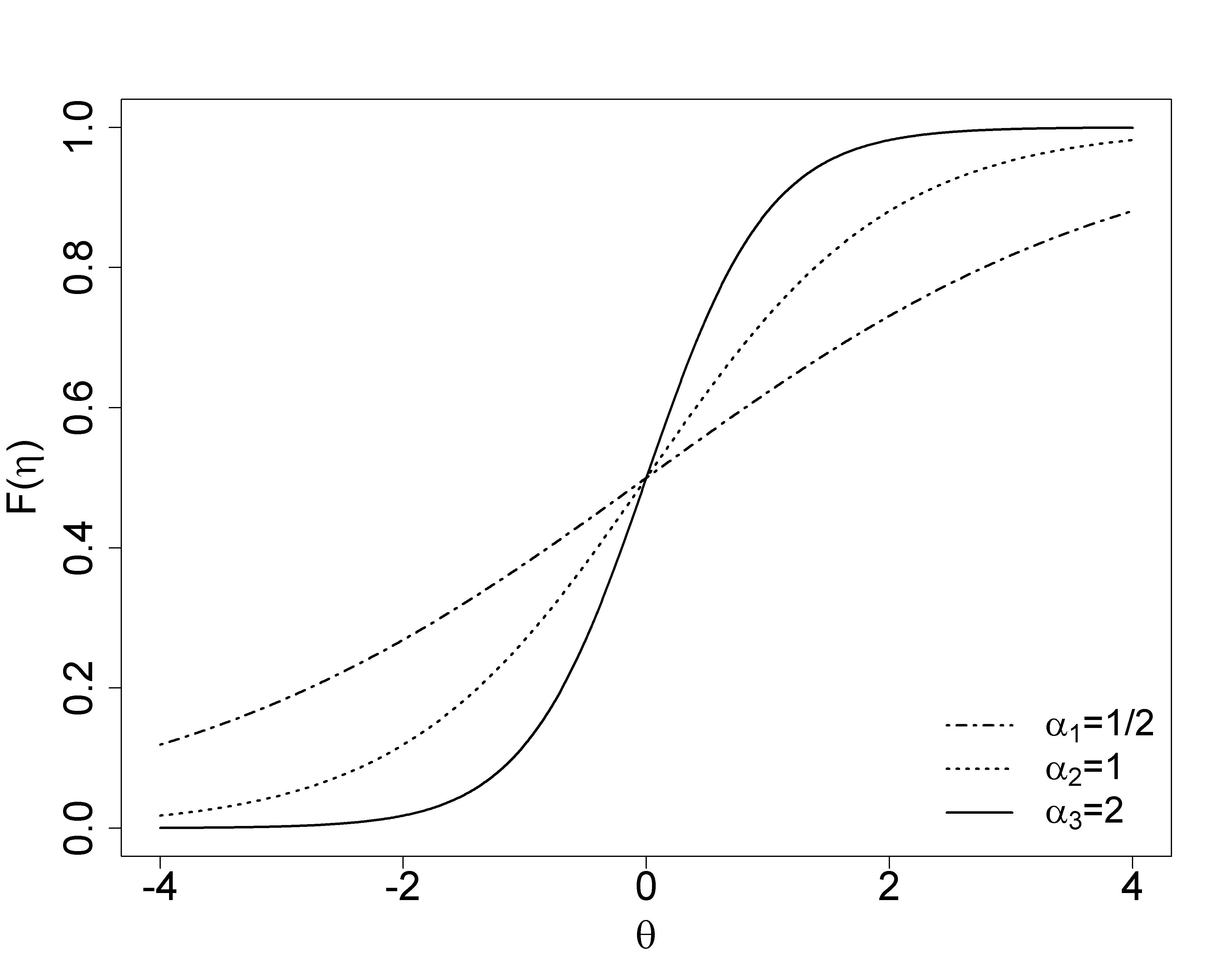}
  \caption{CdF adjustment}
  \label{fig_logit_irt}
\end{subfigure}
\caption{Relationship between the CdF and the IRT parameterization for dichotomous items where $F(\eta)=F(\alpha_j(\theta-\delta_j))$. 
Figure \ref{fig_cdf} presents the different CdF for one item $j$ given $\alpha_{j}=1$ and $\delta_{j}=0$. Figure \ref{fig_logit_irt} presents the logistic CdF adjustment for three items ($j=1,2,3$) with different $\alpha_j$ and $\delta_j=0$, given the linear predictor defined in equation \eqref{eq_classical_eta}. \label{figure1}}
\end{figure}

Figure \ref{fig_cdf} shows different slopes depending on the particular CdF. 
The CdF allows to take into account the influence of linear predictor ($\eta$) change on the response probability evolution. 
In general IRT parameterization (equation \ref{eq_classical_eta}), the slope adjustment is managed by the discrimination parameter. Depending on different discrimination parameter values, Figure \ref{fig_logit_irt} presents the CdF logistic according to the individual latent variable. 
This item parameter has the task of fitting the CdF for each considered item to distinguish more the different response variable.

In the literature, the cumulative model is presented according to the use of several of the previously mentioned CdF \citep{samejima_estimation_1968,fahrmeir_multivariate_2001,liu_mixed-effects_2006}, while the adjacent models are most often presented with the logistic CdF. 
However, for both, we recommend the use of a symmetric CdF. 
As seen in Figure \ref{fig_b}, the IRT parametrization is a subtle and effective way to take into account the multiple item outcome characteristics in GLMM for categorical responses.
In the context of the HRQoL in clinical trial, the HRQoL dimension considers a small set of items which are correlated and measure an unique latent variable. 
The discrimination parameter is routinely not use in this kind of analysis.\\

Relatively to the literature, Table \ref{tab_IRT_GLMM} shows the specification of the famous polytomous IRT models following the different components. For IRT models being within the class of GLMM, we proposed to define them with the four components \textit{(r,F,$Z_q$,$U_r$)}.
The kind of considered location item parameters can be indicated by the index $q$ where $q=1$ for including only difficulty parameters. 
Let $q=2$ for considering the rating scale model \citep{andrich_rating_1978} parameterization where the difficulty parameters are common for all items and one shift parameter is considered for each item. 
Regarding the random part, the number of random effects can be then indicate by the index $r$. 
For the classical IRT parametrization presented in Table \ref{tab_IRT_GLMM}, only one random effect ($r=1$) is taken into account : the capacity parameter $\theta$.
For IRT models including discrimination parameters for each item, we proposed to replace the components $Z$ and $U$ by a component specifying that the predictor is no longer linear (nl), such as \textit{(r,F,nl)}.

\begin{table}[tbh!]
\caption{Specification of the famous IRT model following the components : \textit{(r,F,$Z_p$,$U_r$)} for the GLMM and \textit{(r,F,nl)} for IRT model with no longer linear predictor. Index $p$ denotes the number of kind of item parameters considered in the IRT model and $r$ the number of random effect. }\label{tab_IRT_GLMM}
\begin{center}
\scalebox{1}{
\begin{tabular}{lcc}
 \hline 
 IRT models & $\eta_{im}^{(j)}$ & \textit{(r,F,$Z_p$,$U_r$)} \\ 
 \hline 
 Rating scale model & $\theta_{i}-\left(\delta_{m}+\tau_j\right)$ & \textit{(adjacent,logistic,$Z_2$,$U_1$)} \\ 
 Partial credit model & $\theta_{i}-\delta_{jm}$ & \textit{(adjacent,logistic,$Z_1$,$U_1$)} \\ 
 Sequential Rasch model & $\theta_{i}-\delta_{jm}$ & \textit{(sequential,logistic,$Z_1$,$U_1$)} \\ 
 \hline
 Graded response model & $\alpha_j\left(\theta_{i}-\delta_{jm}\right)$ & \textit{(cumulative,logistic,nl)} \\ 
 Generalized partial credit model & $\alpha_j\left(\theta_{i}-\delta_{jm}\right)$ & \textit{(adjacent,logistic,nl)} \\ 
 \hline
 \end{tabular}  
 }
\end{center}
\end{table}

\section{Simulation study}

In the previous section, we focused on the use of the mixed models for ordinal data analysis and their relevance in the HRQoL analysis in oncology was discussed. Some comparisons studies exist between these different approaches \citep{blanchin_comparison_2011,anota_comparison_2014,barbieri_applying_2015}, mainly on the fixed part of the mixed models
 to identify the trend of latent trait.
\citet{anota_comparison_2014} had shown an equivalent capacity to detect a fixed effect for the LMM and for one of the IRT models. 
Indeed, even if the LMM take into account the HRQoL score, which is a summary variable, this approach is at least equivalent to the IRT models in terms of power.

In this simulation study, the adjacent and cumulative models with the same parameterization of the linear predictor and the logistic CdF were used (as usually in the IRT models).
The aim of the following section is to reinforce these comparisons between the LMM and the IRT models on the random part of the mixed models. The datasets were simulated from an IRT model (adjacent and cumulative models). Regarding the parameterization, two subject-specific random effects $\xi_{i0}$ and $\xi_{i1}$ were considered, respectively associated with the intercept and the slope. Of course, the usefulness of the random effect introduction in the model is strongly associated with the observed data. 
HRQoL is a subjective endpoint, and the individual random effect $\xi_{i0}$ is thus entirely justified. Indeed, it is easy to imagine that each patient has a different level of HRQoL at baseline. The random slope is more questionable, indeed, the assumption that the specific HRQoL evolution of one single patient diverges from the average evolution for the whole population, is less obvious than the previous one. In this section, the capacity of the mixed models to detect the slope random effect was thus studied. No group effect was considered in this simulation study.

\subsubsection*{Design}

We want to study the capacity of each model to detect the random effect $\xi_{i1}$ associated with time (random slope). 
The two subject-specific random effects are considered independent where $\xi_{i0}\sim\mathcal{N}(0,\sigma_0^2)$ and $\xi_{i1}\sim\mathcal{N}(0,\sigma_1^2)$. The following model choice study is performed on the basis of the Bayesian information criteria (BIC) where two models were considered: $\mathcal{M}_2$ with the two random effects \textit{(r,F,$Z_1,U_2$)} and $\mathcal{M}_1$ excluding the random slope \textit{(r,F,$Z_1,U_1$)}. 
For the IRT models, the linear decomposition of the latent trait $\theta_{iv}$ only took into account the time as a fixed effect. The two considered models with proportional design are:
\begin{equation}
  \left.
      \begin{aligned}
\mathcal{M}_2: \theta_{iv} & =  \left(t_v-t_0\right)\beta_1+\xi_{i0}+\left(t_v-t_0\right)\xi_{i1}\\
\mathcal{M}_1: \theta_{iv} & =  \left(t_v-t_0\right)\beta_1+\xi_{i0}
      \end{aligned}
    \right.
    \label{choicemodel_IRT}
\end{equation}

In order to best reflect the EORTC QLQ-C30 questionnaire, the most frequent HRQoL dimension with two items ($j = 1,2$) comprising four response categories ($m\in\left\{0,\ldots,M\right\}$ with $M= 3$), was used to design the simulation study. A sample size of three hundred subjects ($i=1,\ldots,n$ with $n=300$) and eight follow-up time ($v=0,…,7$), as for the trial presented in the previous section, were considered. 
The datasets were simulated from a multinomial distribution. The different response probabilities $\left\{\pi_{ivm}^{(j)}=\Pr\left(Y_{iv}^{(j)}=m\vert \theta_{iv},\delta_j\right)\right\}$ concerning the subject $i$ for item $j$ were determined by equation \eqref{adj_cdf} for the adjacent model and by equation \eqref{cum_cdf} for the cumulative model, given:
\begin{itemize}
\item the item parameters $\left(\delta_{j1},\delta_{j2},\delta_{j3}\right)_{j=1,2}$;
\item the latent trait ($\theta_{iv}$) deduced in accordance with equation \eqref{choicemodel_IRT};
\item the logistic CdF,
$$
F\left(\eta_{ivm}^{(j)}\right)=\frac{\exp\left(\eta_{ivm}^{(j)}\right)}{1+\exp\left(\eta_{ivm}^{(j)}\right)},
$$
where $\eta_{ivm}^{(j)}=\theta_{iv}-\delta_{jm}$.
\end{itemize}

The values of the parameters used were deduced from the pain symptom data of the clinical trial presented in the previous section. We considered two kinds of difficulty parameters: near $\delta^{ne}=(\delta_1^{ne},\delta_2^{ne})$ and far $\delta^{fa}=(\delta_1^{fa},\delta_2^{fa})$.
These parameter values were chosen in order to illustrate several scenarios described in Table \ref{tab_delta}. The different scenarios were due with the different associations between the model used to simulate the data, \textit{(adjacent,logistic,$Z_1,U_r)_{r=1,2}$} or \textit{(cumulative,logistic,$Z_1,U_r)_{r=1,2}$}, and the different considered values of the difficulty parameters. Table \ref{tab_delta} shows the simulated responses expected at baseline ($t=0$). The responses simulated across time depended of the considered coefficient $\beta_1$. 
Each scenario was simulated $N=500$ times.

\begin{table}[tbh!]
\begin{center}
\caption{\label{tab_delta} Values of difficulty parameters used to simulate the data and expected responses at $t_0$ under each studied scenarios.}
\scalebox{0.9}{
\begin{tabular}{lcc}
\hline
& \multicolumn{2}{c}{Difficulty parameters} \\
\cline{2-3}
Models & $\delta_1^{ne}=(-1.6,1,1.45)$ & $\delta_1^{fa}=(-2.1,1,2.75)$ \\ 
$(r,F,Z_1,U_r)_{r=1,2}$ & $\delta_2^{ne}=(-0.8,1.15,1.9)$ & $\delta_2^{fa}=(-1.25,1.4,3.3)$ \\
\hline 
\textit{(adjacent,logistic,$Z_1,U_r)_{r=1,2}$} & balanced responses & focus on center categories (1 and 2) \\
\textit{(cumulative,logistic,$Z_1,U_r)_{r=1,2}$} & focus on extreme categories (0,1 and 3) & balanced responses  \\
\hline
\end{tabular}
}
\end{center}
\end{table}

Concerning the LMM, the scoring procedure proposed by the EORTC was considered \citep{fayers_eortc_2001}, and the score associated with a symptomatic dimension was first calculated using the simulated data.
Let the two simulated ordinal outcomes $y_{iv}^{(1)}$ and $y_{iv}^{(2)}$ concerning the individual $i$ at the visit $v$, the related score was:
\begin{equation}
S_{iv}=\left(\frac{\sum_{j=1}^{J=2}y_{iv}^{(j)}}{2}\right) \frac{100}{M}\nonumber
\label{eq_score}
\end{equation}
Similarly to the parameterization in equation \eqref{choicemodel_IRT}, we took into account the related choice model with:
\begin{equation}
\label{choicemodel_LMM}
  \left.
      \begin{aligned}
\mathcal{M}_2: S_{iv} & = \beta_0^{^{l}} + \left(t_v-t_0\right)\beta^{^{l}}_1+\xi_{i0}+\left(t_v-t_0\right)\xi_{i1} + \varepsilon_{iv}\\
\mathcal{M}_1: S_{iv} & = \beta_0^{^{l}} + \left(t_v-t_0\right)\beta_1^{^{l}}+\xi_{i0} +\varepsilon_{iv}
      \end{aligned}
    \right.\nonumber
\end{equation}
where $\beta_0^{^{l}}$ is the fixed parameter associated with the intercept, the  $\xi^{^{l}}$ are the random effects normally distributed with the mean equals to zero and  $\varepsilon_{iv}\sim\mathcal{N}(0,\sigma^{2}_{\varepsilon})$ the error term.

\subsubsection*{Results}

Table \ref{table_simu1} shows the capacity of the three models (adjacent model, cumulative model and LMM) to detect the random slope given different scenarios (Table \ref{table_recap}). 
When we simulated the data under $\mathcal{M}_2$ according to the random effect variances estimated from real data, each model detected the random slope ($\xi_{i1}$) in $100\%$ of cases whatever the different situations. 
On the contrary, under $\mathcal{M}_1$, the simulated model $\mathcal{M}_1$ was correctly chosen in most cases. 
For all simulations under $\mathcal{M}_1$, the cumulative model seemed to detect the random slope although it was not included in the simulation step. 
Moreover, the IRT model which was not use to simulate the data, wrongly detected this random effect given a negative value of $\beta_1$ and the difficulty parameter coefficients $\delta^{ne}$. This could be explained by the fact that the difficulty parameters were not uniformly separated around zero and also because they were too close. Indeed, given $\beta_1<0$, the probabilities to observe the upper categories were very small over time and under-represented in comparison with the lower categories (as illustrated in Figure \ref{fig_a}). 
In the specific case where $\beta_1=-0.3$, the IRT model which did not simulate the data could not explain the different outcomes only with the fixed effect and the random intercept, and it compensated the lack of information with the random slope. 
We then could expect symmetric results from $\beta_1$ (positive values) considering the opposite sign of the difficulty parameters because of the reversibility property of the IRT models.

On the contrary, the LMM was stable and thus allowed making the good choice of model whatever the $\beta_1$ values and the IRT model used to simulate the data. Concerning the IRT models where only one model out of the two detected the random effect $\xi_{i1}$, the most suitable model seemed the one not detecting this random effect.

\begin{table}[tbh!]
\begin{center}
\caption{\label{table_simu1} Frequency (on $N=500$ datasets) of the $\mathcal{M}_1$ selection according to the BIC, given $t_v=(0,1,2,4,6,8,10,12)$ and $\sigma_0^2=1.5$. For $r=1,2$, the \textit{(adjacent,logistic,$Z_1,U_r)$} models and the \textit{(cumulative,logistic,$Z_1,U_r)$} models are denoted respectively by AM and CM. For the random component, $U_1$ if $\sigma_1^2=0$ and $U_2$ if $\sigma_1^2>0$.} 
\scalebox{0.9}{
\begin{tabular}{cc|c|c|c|c|c|c|c|c|c|c|c|c}
\hline
\hline
\multicolumn{2}{l}{Simulated} & \multicolumn{12}{c}{Scenarios} \\
\multicolumn{2}{l}{Model} & \multicolumn{3}{|c|}{AM using $\delta^{ne}$} & \multicolumn{3}{|c|}{CM using $\delta^{fa}$} & \multicolumn{3}{|c|}{CM using $\delta^{ne}$} & \multicolumn{3}{|c}{AM using $\delta^{fa}$}\\ 
\hline
$\sigma_1^2$ & $\beta_1$ & LMM & AM & CM & LMM & AM & CM & LMM & AM & CM & LMM & AM & CM \\ 
\hline
\hline
$0.2$ &  $-0.3$ & 0 & 0 & 0 & 0 & 0 & 0 &  0 & 0 & 0 & 0 & 0 & 0 \\ 
$0.2$ & $0.3$ & 0 & 0 & 0 & 0 & 0 & 0 & 0 & 0 & 0 & 0 & 0 & 0 \\ 
\hline 
$0$ & $-0.5$ & 97.67 & 99.29 & {56.49} & 100 & 94.63 & 92.98 & 100 & {61.33} & 95.71 & 100 & 99.66 & 89.54 \\ 
$0$ & $-0.3$ & 99.00 & 100 & {33.04} & 100 & 88.63 & 93.30 & 100 & {36.33} & 94.91 & 100 & 100 & 83.33 \\ 
$0$ & $-0.2$ &  100 & 99.62 & {49.28} & 100 & 94.56 & 93.81 & 100 & {71.67} & 95.77 & 100 & 99.64 & 79.02 \\ 
$0$ & $-0.1$ & 98.67 & 95.65 & 94.78 & 100 & 98.65 & 89.62 & 100 & 98.98 & 90.41 & 100 & 100 & 88.10 \\ 
$0$ & $0.0$ & 95.60 & 100 & 94.55 & 99.00 & 99.66 & 91.75 & 99.00 & 99.66 & 89.71 & 97.00 & 99.66 & 94.42 \\ 
$0$ & $0.1$ & 83.00 & 100 & 94.78 & 93.33 & 100 & 92.63 & 97.00 & 100 & 90.91 & 87.33 & 100 & 94.69 \\ 
$0$ & $0.3$ & 98.33 & 99.64 & 90.61 & 100 & 99.64 & 89.05 & 100 & 100 & 93.67 & 100 & 99.65 & 93.78 \\ 
$0$ & $0.5$ & 100 & 100 & 94.29 & 100 & 99.32 & 94.71 & 100 &100 &97.61  & 100 & 100 & 97.19 \\
\hline 
\end{tabular}
}
\end{center}
\end{table}

The capacity of the different models to detect the random slope when its variance value changes, is presented in Table \ref{table_simu2}. 
All models were sensitive to the signal-to-noise ratio. Indeed, the more $\beta_1$ increased, the less the random effect provided information. This was well characterized as the capacity to detect the random effect for greater variances when the signal was strong. 
In this case, the signal provided the essential information explaining the different responses. In the model comparison, the LMM was less sensitive than the IRT models. Indeed, the LMM detected the random slope for a greater variance of this one whatever the $\beta_1$ value. This result was expected because the LMM is based on the HRQoL score which is a summary variable with less information than the raw data. 
Thus, the IRT models are more sensitive in all cases. 
Comparing the two IRT models,
there is a tendency for the random slope model to be preferred under the cumulative model regardless of whether it is the true model model or not.
On the contrary, in the specific case where $\beta_1=-0.3$, the IRT model used to simulate the data was less efficient than the other IRT model which detected a random slope to remedy the lack of information. This was coherent with our previous results shown in Table \ref{table_simu1}. Finally, the more $\beta_1$ was close to zero, the more the models detected the random slope for a low variance.

\begin{table}[tbh!]
\begin{center}
\caption{\label{table_simu2} Frequency (on $N=500$ datasets) of the $\mathcal{M}_2$ selection according to the BIC, given $t_v=(0,1,2,4,6,8,10,12)$ and $\sigma_0^2=1.5$. For the \textit{(adjacent,logistic,$Z_1,U_r)$} models and the \textit{(cumulative,logistic,$Z_1,U_r)$} models are denoted respectively by AM and CM. For the random component, $U_1$ if $\sigma_1^2=0$ and $U_2$ if $\sigma_1^2>0$.}
\scalebox{0.85}{
\begin{tabular}{cc|c|c|c|c|c|c|c|c|c|c|c|c}
\hline
\hline
\multicolumn{2}{l}{Simulated} & \multicolumn{12}{c}{Scenarios} \\
\multicolumn{2}{l}{Model} & \multicolumn{3}{|c|}{AM using $\delta^{ne}$} & \multicolumn{3}{|c|}{CM using $\delta^{fa}$} & \multicolumn{3}{|c|}{CM using $\delta^{ne}$} & \multicolumn{3}{|c}{AM using $\delta^{fa}$}\\ 
\hline
 $\beta_1$ & $\sigma_1^2$ & LMM & AM & CM & LMM & AM & CM & LMM & AM & CM & LMM & AM & CM \\ 
\hline
\hline
& $0.01$ & 0 & 2.33 & 24.92 & 0 & 5.03 & 6.94 & 0 & 2.69  & 3.70 & 0.33 & 6.44 & 24.75 \\ 
& $0.02$ & 0 & 21.40 & 54.67 & 0 & 37.58 & 44.11 & 0 & 17.73 & 18.12 & 0 & 50.00 & 77.00  \\ 
& $0.03$  & 0 & 61.00 & 90.97 & 0 & 75.67 & 80.00 & 0 & 41.33 & 45.58 & 0 & 86.33 & 98.33 \\ 
$1$ & $0.05$ & 0 & 97.67 & 99.66 & 0 & 100 & 100 & 0.33 & 89.00 & 90.00 & 0 & 99.33 & 100 \\  
& $0.2$ &  39.33 & 100 & 100  & 40.67 & 100 & 100 & 10.67 & 100 & 100 & 57.67 & 100 & 100 \\ 
& $0.5$ & 100 & 100 & 100 & 100 & 100 & 100 & 100 & 100 & 100 & 100 & 100 & 100\\ 
\hline 
& $0.002$ & 16.67 & 6.33 & 21.40 &  0 & 2.03 & 3.97 & 0 & 3.06 & 3.94 & 11.00 & 11.04 & 15.25 \\ 
& $0.005$ & 72.33 & 86.33 & 92.67  & 30.67 & 55.33 & 59.00 & 0 & 32.33 & 46.00 & 85.67 & 87.33 & 91.67 \\ 
$0.3$ & $0.008$ & 97.67 & 100 & 100  &  86.00 & 97.3 & 98.00 & 4.00 & 76.33 & 88.33 & 99.33 & 99.67 & 100 \\  
& $0.01$ & 100 & 100 & 100 & 96.33 & 99.67 & 99.33 & 17.33 & 94.00 & 97.00  & 100 & 100 & 100\\ 
& $0.02$ & 100 & 100 & 100 & 100 & 100& 100 & 96.67 & 100 & 100 & 100 & 100 & 100\\ 
\hline
& $0.002$ & 0.67 & 4.36 & 61.43 & 0 & 54.00 & 5.09 & 0 & 93.33 & 1.82 & 0 & 2.07 & 18.57 \\ 
$-0.3$ & $0.005$ & 5.67 & 62.33 & 79.00 & 0 & 95.67 & 40.40 & 0 & 99.67 & 33.22 & 0 & 56.00 & 48.33 \\ 
& $0.008$ & 23.67 & 96.33 & 97.33 & 0 & 100 & 86.67 & 0 & 100 & 82.67 & 1.67 & 96.33 & 86.33 \\  
\hline
\end{tabular}
}
\end{center}
\end{table}

\section{Application to HRQoL data}

In this section, we performed a longitudinal analysis of the HRQoL data from a multicenter randomized phase III clinical trial in first-line metastatic pancreatic cancer patients: PRODIGE4/ACCORD11 \citep{conroy_folfirinox_2011}. 
Three hundred and forty-two patients were randomly assigned to Folfirinox (experimental arm) versus Gemcitabine (control arm) regimens. The detailed inclusion and exclusion criteria, the study design and protocol, the treatment, the compliance to the questionnaires, and the HRQoL analyses have previously been published \citep{conroy_folfirinox_2011,gourgou-bourgade_impact_2012,barbieri_applying_2015}. The patients filled the EORTC QLQ-C30 questionnaire themselves at different follow-up times defined in the protocol: at baseline, day 15, day 30, and at months 2, 4, 6, 8, and 10. The different measuring times reflected the longitudinal aspect of the HRQoL and allowed the assessment of the change of HRQoL for each dimension. 

Analyses were performed using the SAS software (version 9.3) \citep{institute_sas/stat_2011,boeck_explanatory_2004}. For all previous arguments, the cumulative models are preferred for the longitudinal analysis of the HRQoL. There, we considered the \textit{(cumulative,logistic,$Z_1,U_2)$} model with 
to analyze the data.
In HRQoL study in oncology, the analysis is carried out for each HRQoL dimension. Given one HRQoL dimension with few correlated items, the discrimination parameters could be considered equals to one for each item. 
The distinction between the multiple-item responses is only achieved though difficulty parameters (thresholds) \citep{anota_comparison_2014,barbieri_applying_2015}.
For the HRQoL longitudinal analysis with the subject $i\left(i=1,\ldots,342\right)$, the visit $v\left(v=1,\ldots,8\right)$, the item $j$ with $M_j$ response categories, the \textit{(cumulative,logistic,$Z_1,U_2)$} model is defined by:
\begin{equation}
\Pr\left(Y^{(j)}_{iv}\geq m\vert \theta_i\right) = \frac{exp\left(\eta^{(j)}_{ivm} \right)}{1+exp\left(\eta^{(j)}_{ivm} \right)},
\end{equation}
with the following linear predictor is considered in the analyses:
\begin{equation}
  \left\{
      \begin{aligned}
       \eta_{ivm}^{(j)} = & \theta_{iv}-\delta_{jm} \\
	   \theta_{iv} = & g_i\beta_1 + \left(t_v-t_0\right)\beta_2+g_i\left(t_v-t_0\right)\beta_3\\
	   &  +\xi_{i0}+\left(t_v-t_0\right)\xi_{i1}
      \end{aligned}
    \right.
    \label{example}
\end{equation}
where:
\begin{itemize}
\item $\delta_{jm}$ is the difficulty parameter (threshold) associated with the category $m$ of item $j$;
\item $t_v$ is the date of the visit $v$, and $t_0$ is the date of baseline;
\item $g_i=1$ if the patient $i$ belongs the experimental group (Folfirinox), $g_i=0$ if the patient $i$ belongs the control group (Gemcitabine);
\item $\beta_1$ is the effect difference at baseline between Folfirinox and Gemcitabine groups;
\item $\beta_2$ is the slope (evolution) of health-related quality-of-life perception for the Gemcitabine group;
\item $\beta_2+\beta_3$ is the slope (evolution) of health-related quality-of-life perception for the Folfirinox group;
\item $\xi_{i0}$ and $\xi_{i1}$ are respectively the subject-specific random effects associated with the intercept and the slope such as $\left(\xi_{i0},\xi_{i1}\right)'\sim\mathcal{N}\left(\textbf{0},\Sigma\right)$, $\Sigma$ being the unstructured covariance matrix.\\
\end{itemize}

These HRQoL data have been already analyzed with different approaches. 
\citet{gourgou-bourgade_impact_2012} have presented the results using time-to-event models. They concluded for a better HRQoL in the Folfirinox arm than the Gemcitabine arm.
Then, \citet{barbieri_applying_2015} have presented the results through the LMM and the partial credit model extended for the longitudinal analysis \textit{(adjacent,logistic,$Z_1,U_2)$}. 
The conclusions of both mixed models are similar.

For the \textit{(cumulative,logistic,$Z_1,U_2)$} model, Table \ref{tab_estimations} shows the estimations of fixed parameters, their standard deviation and the associated P-value of the Wald test. 
Concerning the functional dimension, we performed a reverse permutation on the functional scales for an intuitive interpretation. 
This allows to consider that an increase of the latent trait $\theta$ is associated with an increase of the functional capacity (improvement of HRQoL) or increase of the symptoms (deterioration of HRQoL). 
For all HRQoL dimensions, there should be no difference at baseline ($\beta_1=0$) in a randomized clinical trial.
However, we observed a significantly difference concerning the diarrhea symptom between the two groups at baseline ($p=0.007^{**}$). It referred to a difference between the two arms at day 15, day 30 (during treatment period) but no necessarily at baseline. Then, the perception of diarrhea symptom remained higher in the Folfirinox arm over time.
This result is expected because the Folfirinox is more toxic than Gemcitabine and is known to cause more diarrhea symptom.

Regarding the others dimensions, the HRQoL changed over time for several dimensions (emotional functioning, pain, insomnia, constipation and appetite loss) returning a significant improvement in terms of HRQoL perception.
Only the pain showed a significantly different evolution between the two arms ($p=0.04$). 
Indeed, the patients receiving the Folfirinox had a perception of pain which decreased significantly more over time than the patients receiving the Gemcitabine.

\begin{table}[tbh!]
\begin{center}
\caption{Estimations of fixed effect parameters $(\beta_p)_{p=1,2,3}$ of the \textit{(cumulative,logistic,$Z_1,U_2)$} model. All HRQoL dimensions of the EORTC QLQ-C30 are considered.\label{tab_estimations}}
\scalebox{0.8}{
\begin{tabular}{llccc}
\hline
\multicolumn{2}{l}{HRQoL}  &  \\
\multicolumn{2}{l}{Dimensions}  & Coefficient & Standard error & \textit{Pvalue} \\
\hline
\hline
\multicolumn{2}{l}{Global Health Status} \\
 & $\beta_2$ &  0.098 & 0.070 & 0.166 \\
 & $\beta_3$ &  0.130 & 0.085 & 0.128 \\
\hline
\multicolumn{2}{l}{Physical functioning} \\
 & $\beta_2$ & -0.150 & 0.077 & 0.051 \\
 & $\beta_3$ & 0.122 & 0.098 & 0.212 \\
\hline
\multicolumn{2}{l}{Role functioning} \\
 & $\beta_2$ &  -0.011 & 0.081 & 0.892 \\
 & $\beta_3$ &  0.157 &  0.103 &  0.131 \\
\hline
\multicolumn{2}{l}{Emotional functioning} \\
 & $\beta_2$ & 0.335 & 0.070 & $<.001^{***}$\\
 & $\beta_3$ & 0.001&   0.086 & 0.992\\
\hline
\multicolumn{2}{l}{Cognitive functioning} \\
 & $\beta_2$ &  -0.002 & 0.054 & 0.972\\
 & $\beta_3$ &  0.088 & 0.067 & 0.189\\
\hline
\multicolumn{2}{l}{Social functioning} \\
 & $\beta_2$ &  0.010 & 0.073 & 0.888\\
 & $\beta_3$ &  0.116 & 0.093 & 0.211\\
\hline
\hline
\multicolumn{2}{l}{Fatigue} \\
 & $\beta_2$ & -0.087 & 0.085 & 0.308  \\
 & $\beta_3$ & -0.033 & 0.107 & 0.761  \\
\hline
\multicolumn{2}{l}{Nausea and vomiting} \\
& $\beta_2$  & -0.052 & 0.060 & 0.393 \\
 & $\beta_3$ & -0.069 & 0.072 & 0.336  \\
\hline
\multicolumn{2}{l}{Pain} \\
 & $\beta_2$ & -0.330 & 0.076 & $<.001^{***}$ \\
 & $\beta_3$ &-0.188 & 0.092 & 0.040$^{*}$ \\
\hline
\multicolumn{2}{l}{Dyspnea} \\
 & $\beta_2$ & -0.060 & 0.075 & 0.420\\
 & $\beta_3$ & -0.093 & 0.088 & 0.295\\
\hline
\multicolumn{2}{l}{Insomnia} \\
 & $\beta_2$ & -0.359 & 0.080 & $<.001^{***}$ \\
 & $\beta_3$ &  0.046 & 0.083 & 0.627 \\
\hline
\multicolumn{2}{l}{Appetite loss} \\
 & $\beta_2$ & -0.354 & 0.072 & $<.001^{***}$ \\
 & $\beta_3$ & -0.026  & 0.080  &  0.747\\
\hline
\multicolumn{2}{l}{Constipation} \\
 & $\beta_2$ &   -0.325 & 0.077 & $<.001^{***}$\\
 & $\beta_3$ &   0.003 & 0.083  & 0.974\\
\hline
\multicolumn{2}{l}{Diarrhea} \\
 & $\beta_1$ &  0.739 & 0.272 & 0.007$^{**}$\\
 & $\beta_2$ &  0.018 & 0.067 & 0.792 \\
 & $\beta_3$ & -0.026 & 0.076 & 0.786\\
\hline
\multicolumn{2}{l}{Financial difficulties} \\
 & $\beta_2$  & -0.522 & 0.282 & 0.066 \\
 & $\beta_3$  &  0.302 & 0.208 & 0.146 \\
\hline
\end{tabular} 
}
\end{center}
\end{table}

\begin{figure}[thb!]
\centering
\begin{subfigure}{0.48\textwidth}
  \centering
  \includegraphics[width=\textwidth]{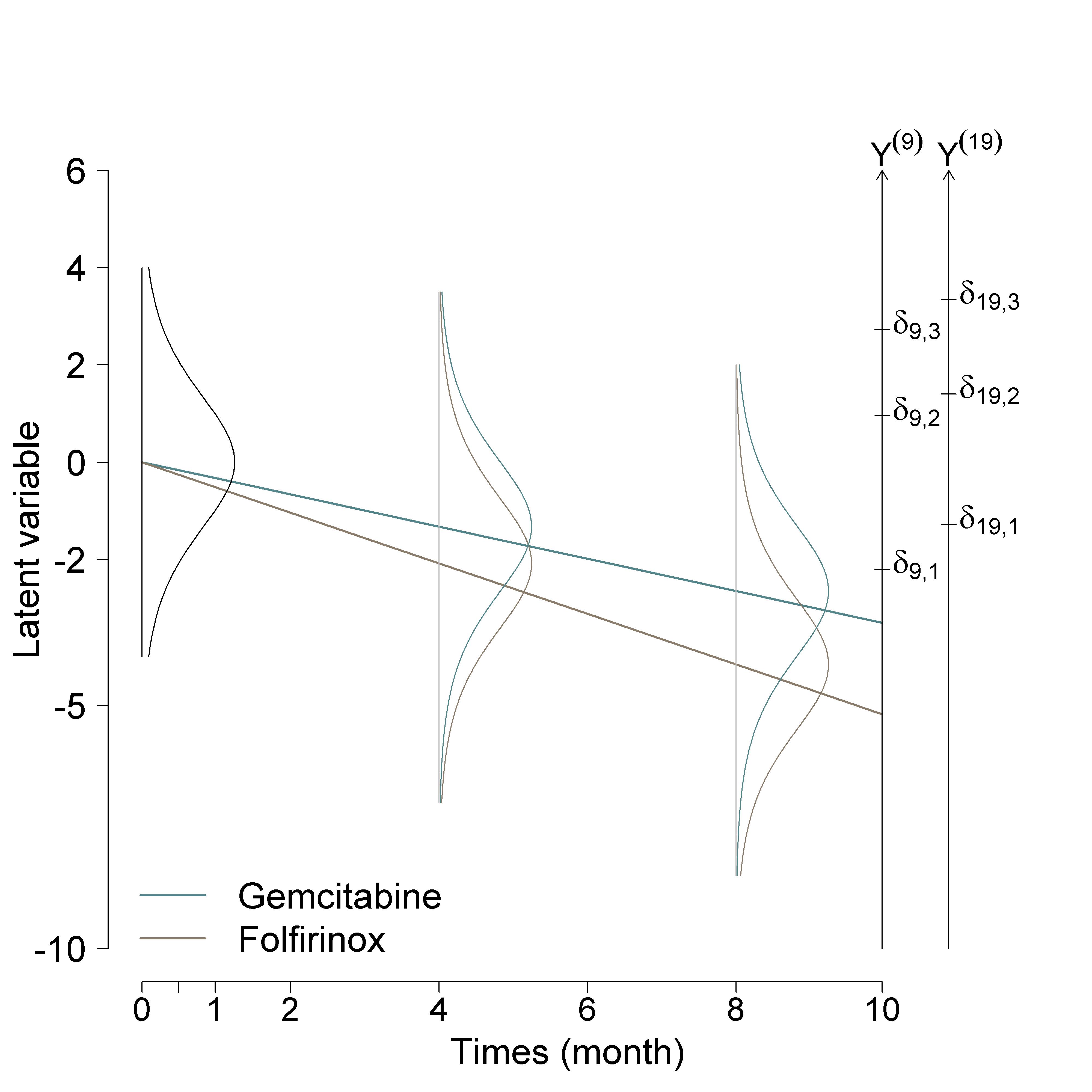}
  \caption{Pain evolution}
  \label{fig_a}
\end{subfigure}
     ~ 
\begin{subfigure}{0.48\textwidth}
  \centering
  \includegraphics[width=\textwidth]{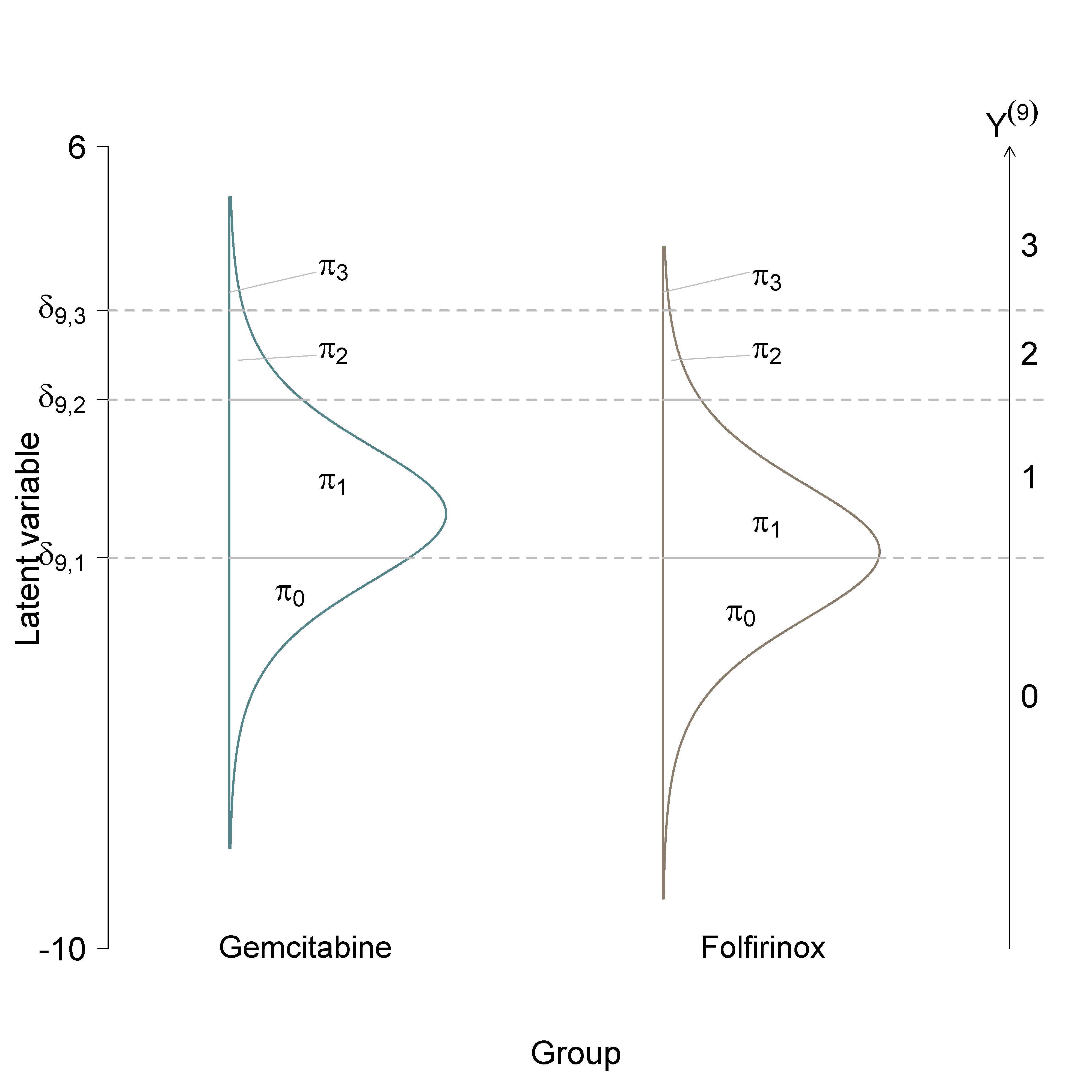}
  \caption{Outcomes at $t_4$}
  \label{fig_b}
\end{subfigure}
\caption{Interpretation of the \textit{(cumulative,logistic,$Z_1,U_2)$} model through its underlying latent variable concerning the pain symptom (including the items 9 and 19). The estimated difficulty parameter for the item 9 are $\delta_{9,1}=-2.1$, $\delta_{9,2}=1$ and $\delta_{9,3}=2.75$, and for the item 19 : $\delta_{19,1}=-1.28$, $\delta_{19,2}=1.40$ and $\delta_{19,3}=3.34$. Figure \ref{fig_a}: the different HRQoL evolution of the latent variable and response variable $(Y^{(j)})_{j=9,19}$ between the two groups. Figure \ref{fig_b}: the different proportions ($\pi_m$) of different response categories of $Y^{(9)}$ between the two groups four month after the Baseline ($t_4$).\label{figure2}}
\end{figure}

Its interpretability and intuitive illustration is one of the many advantages of the cumulative models.
The constraints on the item parameter in these models allow an interpretation through the latent variable (e.g. comparing the proportions of the response categories for one specific item over time or between different groups during a fixed time). 
Figure \ref{figure2} presents the evolution concerning the probability of response either over time (Figure \ref{fig_a}) or between group (Figure \ref{fig_b}). 
This example is illustrated through the first item of the pain symptom of the clinical trial previously presented. 
The probability ($\pi_m$) for a patient to response the category $m$ corresponds to the area under the curve delimited by the horizontal lines. 
For both groups, Figure \ref{fig_a} shows that the probability to choose the categories 2 or 3 decreased over time while the probability to choose the category 0 increased. At baseline, the response proportion for the categories 0, 1, 2 and 3 were respectively $\pi_0=0.10$, $\pi_1=0.62$, $\pi_2=0.22$ and $\pi_3=0.06$ for each group. 
Then, the evolution of the proportions showed a decrease of the level of pain between the baseline and the 4-month visit, and, finally, a decrease of the latent trait over time. 
Likewise, Figure \ref{fig_b} shows the different response proportions between the two groups at four months. 
In the control group, the proportions were $\pi_0=0.29$, $\pi_1=0.61$, $\pi_2=0.08$ and $\pi_3=0.02$ for the categories 0, 1, 2 and 3. In the experimental group, they were $\pi_0=0.47$, $\pi_1=0.48$, $\pi_2=0.04$ and $\pi_3=0.01$. 
The probability to response category 3 was the lowest whatever the group, but was even less likely for patients in experimental group than in control group. In contrary, the probability to response category 0 was more likely in experimental group than in control group. The observed gab corresponded to the difference between the two linear predictors associated with each group only four months after the baseline. One of the interests of this illustration concerns the clinical interpretation. The IRT models thus offer a complete analysis: the general analysis of a HRQoL dimension and the specific analysis for each item \citep{edelen_applying_2007}.

\section{Discussion}

We have explored the suitable mixed models for the longitudinal analysis of the HRQoL in oncology. 
This data coming from questionnaires through Likert scales, we focused on regression models for ordinal data. 
These models have been specified with three components, the linear predictor parameterization, the ratio of probabilities and the CdF \citep{peyhardi_new_2015}. 
In oncology, the analysis being performed on multiple-item measurements associated with one HRQoL dimension \citep{fayers_eortc_2001}, the specific IRT parameterization of the linear predictor is thus used. The item parameters allow to distinguish the outcomes from different items which measure an unique unidimensional latent variable. 
This latent variable was decomposed linearly to take into account the different covariates in the fixed part of the model and to incorporate subject-specific random effects. 
The analysis with IRT models is the richer because they are based on raw data \citep{gorter_why_2015}. 
The analysis can be made on one specific item through the item parameters or on the whole HRQoL dimension \citep{edelen_applying_2007}. 
Indeed, these models take into consideration all available information from the data, it is why the use of this kind of model is more and more studied \citep{gorter_why_2015}.

Then, concerning the choice of the model family, the cumulative and adjacent models are preferred. 
From the ratios of probabilities which characterize them and the symmetric CdF, the practical property of the invariant seems important to interpret the results. 
The cumulative models also assume an underlying continuous latent variable that is associated with a linear mixed regression model \citep{mccullagh_regression_1980,hedeker_random-effects_1994}. This allows a better interpretation and illustration of the results such as the easy analysis of the evolution of the response proportions of the different categories over time or between groups, given one item. 
The adjacent models show the advantage not to have any constraint for the model estimation. 
These models can thus be preferred when the regression is performed on the item part of the linear predictor, given non-proportional design.
Finally, the choice of the CdF essentially depends on the observed data and properties which interest the users.
These IRT models are reversible only if the CdF is symmetric.
Thu, the use a commonly symmetric CdF is preferred (the logistic and the Gaussian distributions).
From the conceptual IRT model selection, the cumulative model seems the most suitable given its advantages for the longitudinal analysis of HRQoL in cancer clinical trials. 

The simulation study showed that the IRT model capacity to detect the random effect was better than the LMM currently used. This result seems natural because the LMM is based on the study of a summary variable with less information. Thus, the variability from data is also reduced. Of course, the usefulness of the random effect introduction in the model is strongly associated with the observed data.
Moreover, the more the difficulty parameters were distinct and the influence of covariates was stronger, the less the random effect provided information. All these results confirmed that the IRT models allow a more detailed analysis to interpret the results from a specific dimension or item. Whatever the IRT model used to generate the data, the LMM remained competitive through these simulations. However, the IRT model that did not generate data, seemed more sensitive to the random slope than the other IRT model used to simulate the dataset. Indeed, in some cases, it tended to detect the random slope while it did not exist. 
In case where one of the two models detects the random slope, the use of the model not detecting the effect as it is, seems the most appropriated concerning a data-driven choice. 
However, we recommend to use only one kind of model (with same components discussed previously) allowing to make the results comparable across HRQoL dimensions.

An aspect that remains to be discussed is the multidimensional aspect of HRQoL. Nowadays in oncology, the different dimensions are analyzed independently of one another, and this causes the problem of multiple tests.
An approach to consider the all HRQoL dimensions would be the use of structural equation modeling. 
This would allow to show the influence of each HRQoL dimension through some factors to explain the global HRQoL and potential structural links between the latent variables.

\section*{Funding}
This study was supported by a grant from the French Public Health Research Institute ({\texttt{www.iresp.net}}) under the 2012 call for projects as part of the 2009-2013 Cancer Plan. 

\section*{Acknowledgements}
We thank Dr. H\'{e}l\`{e}ne de Forges for her editorial assistance and UNICANCER for the data from PRODIGE4 / ACCORD11 clinical trial which is used in this paper.

\bibliographystyle{apalike}
\bibliography{references}

\end{document}